\documentclass[final,5p,times,twocolumn]{elsarticle}

\usepackage{geometry}
\usepackage{amsmath,bm}
\DeclareMathOperator\arctanh{arctanh}

\biboptions{numbers,sort&compress}

\usepackage{lipsum}
\usepackage{flushend}
\usepackage{float}
\usepackage{siunitx}
\usepackage{amsfonts}
\usepackage{enumitem}
\usepackage[tableposition=top]{caption}
\usepackage{subfig}
\usepackage{makecell}
\usepackage{xurl}
\usepackage[usenames,dvipsnames]{color}
\usepackage{gensymb}
\usepackage{booktabs}

\usepackage{caption}
\usepackage{subcaption}

\usepackage{threeparttable}

\usepackage{listings}
\usepackage{xcolor}

\definecolor{codegreen}{rgb}{0,0.6,0}
\definecolor{codegray}{rgb}{0.5,0.5,0.5}
\definecolor{codepurple}{rgb}{0.58,0,0.82}
\definecolor{backcolour}{rgb}{0.95,0.95,0.92}

\lstdefinestyle{mystyle}{
    backgroundcolor=\color{backcolour},   
    commentstyle=\color{codegreen},
    keywordstyle=\color{magenta},
    numberstyle=\tiny\color{codegray},
    stringstyle=\color{codepurple},
    basicstyle=\ttfamily\footnotesize,
    breakatwhitespace=false,         
    breaklines=true,                 
    captionpos=b,                    
    keepspaces=true,                 
    showspaces=false,                
    showstringspaces=false,
    showtabs=false,                  
    tabsize=2
}

\newcommand{\Cs}[2]{#1_\mathrm{#2}}

\begin{document}
\sisetup{separate-uncertainty=true}

\begin{frontmatter}
\author[soheil]{Soheil Solhjoo\corref{cor1}}
\cortext[cor1]{soheil@solhjoo.com; s.solhjoo@rug.nl}
\affiliation[soheil]{organization={ENTEG, Faculty of Science and Engineering, University of Groningen},
            addressline={Nijenborgh 4}, 
            city={Groningen},
            postcode={9747 AG}, 
            country={the Netherlands}}

\title{A short note on approximating the critical strain\\for the onset of dynamic recrystallization}

\begin{abstract}
This note provides a MATLAB code to determine the critical strain associated with the onset of dynamic recrystallization. The code takes a closed-form constitutive model and derives the critical strain by solving $\partial^2 \theta / \partial \sigma^2 = 0$. Moreover, several models that could be used for this purpose are studied.
\end{abstract}

\end{frontmatter}


\section{Introduction}
Recently, I discussed that neural networks may provide the most flexible and accurate phenomenological estimators for stress-strain curves \cite{solhjoo2022revisiting}, at least within the ranges that experimental data available. However, there are cases that an analytical model cannot simply be replaced by neural networks; for example, for estimating the plastic deformation activation energy \cite{solhjoo2022revisiting} one may use Sellars and Tegart's hyperbolic sine constitutive model, or estimating the critical strain for the onset of dynamic recrystallization (DRX) using the method of Poliak and Jonas \cite{poliak1996one} requires smooth constitutive models, e.g., \cite{ebrahimi2007characteristic,solhjoo2010determination,solhjoo2012determination,solhjoo2014determination,chen2014mathematical}, which is not directly achievable by neural networks. For a recent review, see, e.g.,  \cite{varela2020critical}. The current short note presents an overview of the models, provides a general method to obtain their corresponding solutions, and compares them. Moreover, a MATLAB code written to perform the calculations is described in this paper, which can be used as a general tool for similar constitutive models.

\section{The second derivative approach}
There are various methods for identifying the associated strain to the onset of DRX, including metallographic procedures and theoretical methods. The latter is mainly based on the works of Poliak and Jonas \cite{poliak1996one}, suggesting that the onset of dynamic recrystallization can be identified from the inflection point of a $\theta-\sigma$ plot, with $\theta=\partial\sigma/\partial\varepsilon$ being the work-hardening rate, where $\sigma$ is stress and $\varepsilon$ is strain. The models describing stress have a general form of:
\begin{equation} \label{eq:stress_general}
    \sigma=f(\varepsilon, \Cs{\varepsilon}{p}, \sigma_0, \Cs{\sigma}{p},v)
\end{equation}
where $v$ is a set of constants, and subscripts 0 and p refer to $\varepsilon=0$ and the maximum stress, respectively. For such a closed-form solution of $\sigma$, the inflection point can be identified from:
\begin{equation}
    \label{eq:second_derivative}
    \frac{\partial}{\partial\sigma} \left( \frac{\partial\theta}{\partial\sigma} \right) = \frac{\partial^2\theta}{\partial\sigma^2} = 0.
\end{equation}

Calculating the second derivative of the work-hardening rate with respect to stress may be complex. While it can be done by hand, it is a challenging task to write code for. Instead, one can simplify the problem to code using the chain rule
as follows:
\begin{equation}
    \label{eq:second_derivative_chain}
    \frac{\partial}{\partial\varepsilon} \left( \frac{\partial\theta}{\partial\varepsilon} \frac{\partial\varepsilon}{\partial\sigma} \right) \frac{\partial\varepsilon}{\partial\sigma} = 0.
\end{equation}
In equation \ref{eq:second_derivative_chain}, $\partial\varepsilon/\partial\sigma$ can be replaced by $\theta^{-1}$. Moreover, assuming that the solution of equation \ref{eq:second_derivative_chain} cannot be obtained from $\theta^{-1}=0$, it can be further simplified as follows:
\begin{equation}
    \frac{\partial}{\partial\varepsilon} \left( \frac{\partial\theta}{\partial\varepsilon} / \theta \right) = 0.
\end{equation}

\section{MATLAB code} \label{s:code}
To derive a closed-form for the critical strain that solves equation \ref{eq:second_derivative_chain}, we need to perform symbolic calculations. The steps for doing the task are straightforward:
\begin{enumerate}
  \item Define the symbols associated with the parameters that contribute to stress
  \item Define stress
  \item Calculate the work-hardening rate ($\theta=\partial\sigma/\partial\varepsilon$)
  \item Calculate $\theta^{'}_\varepsilon=\partial\theta/\partial\varepsilon$
  \item Calculate $\theta^{'}_\sigma = \theta^{'}_\varepsilon / \theta$
  \item Calculate $\theta^{''}_\varepsilon=\partial\theta^{'}_\sigma/\partial\varepsilon$
  \item Solve $\theta^{''}_\varepsilon = 0$ to obtain $\Cs{\varepsilon}{c}$ \label{step:solution}
  \item Define $\Cs{R}{c} = \Cs{\varepsilon}{c} / \Cs{\varepsilon}{p}$
\end{enumerate}
Probably step \ref{step:solution} results in several solutions of the critical strain, where the definition of $\Cs{R}{c}$ becomes handy, as the acceptable solution must satisfy the following conditions: $\Cs{R}{c} \in \mathbb{R}$ and $0 \leq \Cs{R}{c} \leq 1$.

The following is a MATLAB (MathWorks) code corresponding to these steps.
\begin{enumerate}
  \item
    \begin{lstlisting}[language=Octave]
syms e e_p s_p v \end{lstlisting}
where `e' is strain, `s' is stress, `\_p' indicates the values corresponding to the peak stress, and `v' stands for the constant(s) of the model.

  \item This step depends on the selected model. See section \ref{s:models} for the details.
  
  \item \begin{lstlisting}[language=Octave]
theta = diff(sigma, e); \end{lstlisting}
  
  \item \begin{lstlisting}[language=Octave]
theta_1_e = diff(theta, e); \end{lstlisting}
  
  \item \begin{lstlisting}[language=Octave]
theta_1_s = theta_1_e / theta; \end{lstlisting}
  
  \item \begin{lstlisting}[language=Octave]
theta_2_e = diff(theta_1_s, e); \end{lstlisting}
  
  \item \begin{lstlisting}[language=Octave]
e_c = solve(theta_2_e, e); \end{lstlisting}
  
  \item \begin{lstlisting}[language=Octave]
r = e_c / e_p; \end{lstlisting}
\end{enumerate}

\section{Models} \label{s:models}
This section describes several models that are explicitly developed or used for approximating the critical strain associated with the onset of dynamic recrystallization, and for each one, the founding constitutive model, the MATLAB code associated with steps 1 and 2, and the solution for $\Cs{R}{c}$ are mentioned. Some of these models have been modified to extend their flexibility mainly to incorporate the initial stress values; however, only their original versions are presented here. Moreover, the formulae for other values, such as $\theta$, $\theta^{'}_\varepsilon$, and so on, are not presented here; however, they can be easily obtained via the provided code.

\paragraph{\textbf{ES-7}}
The first model presented here is proposed by Ebrahimi and Solhjoo \cite{ebrahimi2007characteristic}. They worked on the constitutive model of Cingara and McQueen \cite{cingara1992new}, which describes stress as the following:
\begin{equation}
    \sigma = \Cs{\sigma}{p} \left( \frac{\varepsilon}{\Cs{\varepsilon}{p}} \exp\left( 1-\frac{\varepsilon}{\Cs{\varepsilon}{p}} \right) \right)^v,
\end{equation}
which can be defined in the MATLAB code using the following line for its second step:
\begin{enumerate}
  \item[] \begin{lstlisting}[language=Octave]
s = s_p * ((e/e_p)*exp(1-e/e_p))^v \end{lstlisting}
\end{enumerate}
This model results in
\begin{equation}
    \Cs{R}{c} = \frac{\sqrt{1-v}+v-1}{v},
\end{equation}
with $\lim_{v\to 0} \Cs{R}{c} = 0.5$.

\paragraph{\textbf{S-10}}
This model assumes a linear relationship between the work-hardening rate and strain, e.g., $\theta=a\varepsilon+b$, with $a$ and $b$ being arbitrary fitting parameters. Integrating it for certain limits \cite{solhjoo2017two} results in \cite{solhjoo2010determination}:
\begin{equation}
    \sigma = \Cs{\sigma}{p} \left( \frac{\varepsilon}{\Cs{\varepsilon}{p}} \left( 2-\frac{\varepsilon}{\Cs{\varepsilon}{p}} \right) \right)^v,
\end{equation}
which results in \cite{varela2020critical}
\begin{equation}
    \Cs{R}{c} = 1 - \left( 2-v+\sqrt{(1-v)(5-v)} \right)^{-1/2}.
\end{equation}
S-10 can be defined in the MATLAB code using the following line:
\begin{enumerate}
  \item[] \begin{lstlisting}[language=Octave]
s = s_p * ((e/e_p)*(2-e/e_p))^v \end{lstlisting}
\end{enumerate}

\paragraph{\textbf{S-12}}
This model assumes the following constitutive model \cite{solhjoo2012determination}:
\begin{equation}
    \sigma = \Cs{\sigma}{p} \sin{ \left( \frac{\pi\varepsilon}{2\Cs{\varepsilon}{p}}\right)}^v,
\end{equation}
resulting in
\begin{equation}
    \Cs{R}{c} = \frac{2}{\pi}\arctan{\left( \sqrt{1-v} \right)}.
\end{equation}
The following line defines the stress of this model.
\begin{enumerate}
  \item[] \begin{lstlisting}[language=Octave]
s = s_p * sin(2/pi*e/e_p)^v \end{lstlisting}
\end{enumerate}

\paragraph{\textbf{S-14}}
This model describes stress as
\begin{equation} \label{eq:s-14}
    \sigma = \sigma_0 +(\Cs{\sigma}{p}-\sigma_0)\tanh{\left( v_1\frac{\varepsilon}{\Cs{\varepsilon}{p}} \right)^{v_2}},
\end{equation}
resulting in
\begin{equation}
    \Cs{R}{c} = \frac{1}{v_1}\arctanh{\left(\sqrt{ \frac{1-v_2}{1+v_2}} \right)}.
\end{equation}
This model can be defined using the following lines for the two steps of the MATLAB code.
\begin{enumerate}
    \item \begin{lstlisting}[language=Octave]
syms e e_p s_p s_0 v1 v2 \end{lstlisting}

    \item \begin{lstlisting}[language=Octave]
s = s_0 + (s_p - s_0) * tanh(v1*e/e_p)^v2 \end{lstlisting}
\end{enumerate}

\paragraph{\textbf{CFC-14}}
The last model to be mentioned in this work is proposed by Chen and co-workers \cite{chen2014mathematical}, which describes stress as:
\begin{equation}
    \sigma = \Cs{\sigma}{p} \left(1 - \exp{\left( v_1  \left(\frac{\varepsilon}{\Cs{\varepsilon}{p}} \right) ^ {v_2} \right)} \right),
\end{equation}
which results in:
\begin{equation} \label{eq:rc_cfc14_original}
    \Cs{R}{c} = \exp{\left( \frac{2\pi k i}{v_2} \right)} \left(v_1 v_2 \right)^{-1/v_2},
\end{equation}
where $i=\sqrt{-1}$ and $k$ is an arbitrary number; see section \ref{s:boundaries} for further details. This model can be defined in the MATLAB code using the following lines for the first two steps:
\begin{enumerate}
    \item \begin{lstlisting}[language=Octave]
syms e e_p s_p v1 v2 \end{lstlisting}

    \item \begin{lstlisting}[language=Octave]
s = s_p * (1 - exp(v1*(e/e_p)^v2)) \end{lstlisting}
\end{enumerate}

\section{Boundaries of the models} \label{s:boundaries}
\paragraph{\textbf{ES-7, S10, and S-12}}
The conditions on $\Cs{R}{c}$ suggest $0 < v \leq 1$ for all these models, resulting in a maximum $\Cs{R}{c}$ of $0.5$ for ES-7 and S-12, and $1-\sqrt{\sqrt{5}-2} \approx 0.514$ for S-10.

\paragraph{\textbf{S-14}}
The condition of $\Cs{R}{c} \in \mathbb{R}$ implies $0 < v_2 \leq 1$ for S-14 \cite{varela2020critical} and $0<v_1$, but no upper limit can be imposed on $v_1$, meaning that $\Cs{R}{c}$ can vary with no upper limit; therefore, it is crucial to identify the values of $v_1$ and $v_2$ such that they do not cause a violation to the $\Cs{R}{c}$'s conditions.

\paragraph{\textbf{CFC-14}}
For CFC-14, assuming $k=0$ reduces equation \ref{eq:rc_cfc14_original} to $\Cs{R}{c}=\left(v_1 v_2 \right)^{-1/v_2}$ \cite{chen2014mathematical}; however, studying stresses of CFC-14 suggests $v_1 < 0$ and $0 < v_2 \leq 1$, which results in $v_1 v_2 < 0$ and consequently $\Cs{R}{c} \in \mathbb{Z}$, e.g., a not-acceptable complex number. As a remedy, its absolute value can be evaluated instead:
\begin{equation} \label{eq:rc_cfc14_modified}
    \Cs{R}{c} = \left| \exp{\left( \frac{2\pi k i}{v_2} \right)} \left(v_1 v_2 \right)^{-1/v_2} \right| = \left| \left(v_1 v_2 \right)^{-1/v_2} \right| = \left(-v_1 v_2 \right)^{-1/v_2}.
\end{equation}
It should be noted that $\left| \exp{\left( 2\pi k i/v_2 \right)} \right| = 1$, regardless of the assigned value for $k$.

\section{Comparison}
\begin{figure}[t]
    \centering
    \includegraphics[width=\columnwidth]{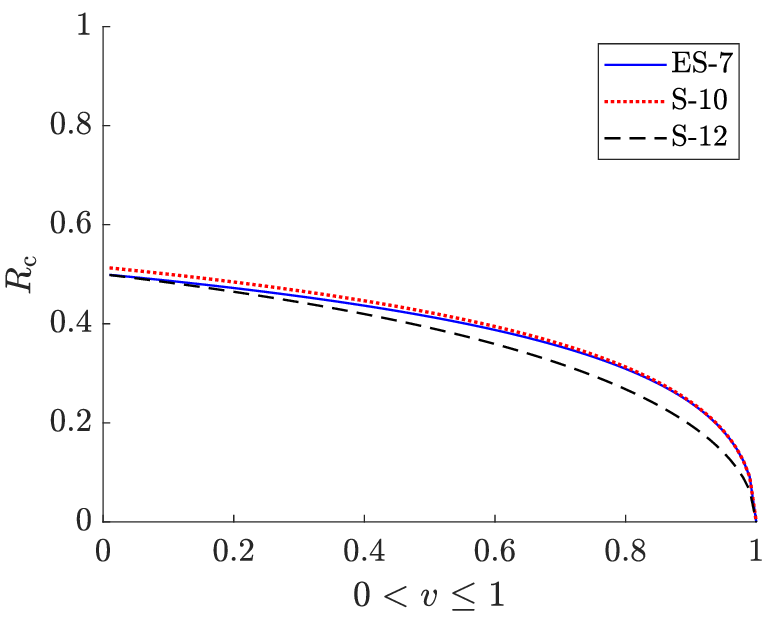}
    \caption{$\Cs{R}{c}$ as a function of $V$ for models ES-7, S-10, and S-12.}
    \label{fig:rc_one_v}
\end{figure}

\begin{figure}[t]
    \centering
    \includegraphics[width=\columnwidth]{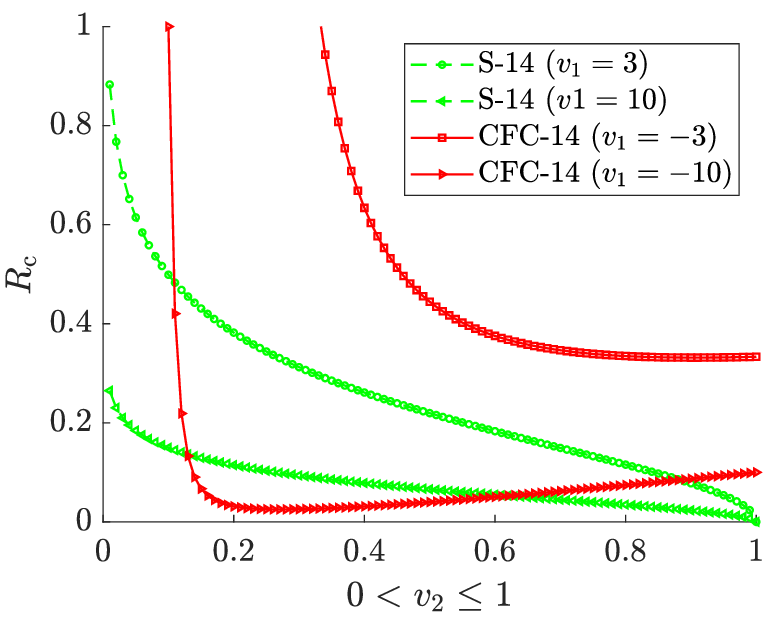}
    \caption{$\Cs{R}{c}$ as a function of $V_1$ and $v_2$ for models ES-14 and CFC-14. For each model, two separate values of $v_1$ are examined for the demonstration.}
    \label{fig:rc_two_vs}
\end{figure}

Figure \ref{fig:rc_one_v} shows the shape of $\Cs{R}{c}$ for the first three models with only one variable, revealing they do not significantly differ from each other. Figure \ref{fig:rc_two_vs} shows the shape of $\Cs{R}{c}$ for S-14 and CFC-14 models with two controlling parameters. The results show that these two models exhibit different behavior than the previous ones with only one variable, for they can cover a broader range of $\Cs{R}{c}$. The behavior of S-14 indicates ease in finding parameters that can satisfy any value of $\Cs{R}{c}$. However, CFC-14 demonstrates an essentially different behavior compared to all other models, indicating difficulties in identifying a pair of $v_1$ and $v_2$ to reach some low values of $\Cs{R}{c}$.


\section{Summary and conclusions}
In this short note, several constitutive models are investigated for their ability to predict the critical strain for the onset of dynamic recrystallization. A simple MATLAB code is proposed to calculate $\Cs{\varepsilon}{c}$ by solving \[\frac{\partial}{\partial\varepsilon} \left( \frac{\partial\theta}{\partial\varepsilon} / \theta \right) = 0.\] Using the code, the closed-form solutions for the normalized critical strain, i.e., $\Cs{R}{c}=\Cs{\varepsilon}{c}/\Cs{\varepsilon}{p}$, are obtained, investigated for their boundaries, and compared.
The results from the behavior of $\Cs{R}{c}$ suggest that, between the studied ones, S-14 (equation \ref{eq:s-14}) is the most suitable model for the intended purpose, which is also the only one (in its original form) that can handle initial stress ($\sigma_0$).


\bibliographystyle{elsarticle-num}
\bibliography{cas-refs}
\end{document}